\documentclass[aps,showpacs,showkeys]{revtex4}
\usepackage{amssymb,amsmath}
\usepackage[dvips]{graphicx}

\newcommand{\bd}[1]{\mbox{\boldmath$#1$}}

\begin{document}

\title{Collective phase dynamics of globally coupled oscillators: \\ Noise-induced anti-phase synchronization}

\author{Yoji Kawamura}
\email[Corresponding author: ]{ykawamura@jamstec.go.jp}
\affiliation{Institute for Research on Earth Evolution,
Japan Agency for Marine-Earth Science and Technology, Yokohama 236-0001, Japan}



\date{December 26, 2013}   



\begin{abstract}
  We formulate a theory for the collective phase description
  of globally coupled noisy limit-cycle oscillators exhibiting macroscopic rhythms.
  Collective phase equations describing such macroscopic rhythms
  are derived by means of a two-step phase reduction.
  The collective phase sensitivity and collective phase coupling functions,
  which quantitatively characterize the macroscopic rhythms,
  are illustrated using three representative models of limit-cycle oscillators.
  As an important result of the theory,
  we demonstrate noise-induced anti-phase synchronization between macroscopic rhythms
  by direct numerical simulations of the three models.
\end{abstract}


\maketitle

\begin{quotation}
{ \noindent
  {\bf Highlights:} \\
  We study globally coupled noisy limit-cycle oscillators showing collective rhythms.   \\
  We formulate the collective phase description of globally coupled noisy oscillators.  \\
  We derive the collective phase equation by means of a two-step phase reduction.       \\
  We illustrate collective phase sensitivity and collective phase coupling functions.   \\
  We demonstrate noise-induced anti-phase synchronization between collective rhythms.   \\
}
\end{quotation}

\section{Introduction} \label{sec:1}

Synchronization phenomena are ubiquitous in systems of coupled limit-cycle oscillators
\cite{ref:winfree80,ref:kuramoto84,ref:pikovsky01,ref:strogatz03,ref:manrubia04,ref:osipov07,ref:mikhailov13}.
The phase reduction method for weakly coupled limit-cycle oscillators
has been fully developed and successfully applied to analyze synchronization properties
\cite{ref:winfree80,ref:kuramoto84,ref:pikovsky01,
  ref:ermentrout96,ref:brown04,ref:hoppensteadt97,ref:izhikevich07,ref:ermentrout10,ref:schultheiss12}.
Collective synchronization in systems of globally coupled phase oscillators
has been particularly investigated~\cite{ref:strogatz00,ref:acebron05,ref:boccaletti06,ref:arenas08,ref:barrat08}.
Such studies consider two typical situations:
the noiseless nonidentical case
\cite{ref:kuramoto84,ref:strogatz00,
  ref:kuramoto75,ref:sakaguchi86,ref:strogatz91,ref:crawford95,ref:crawford99,ref:mirollo07,ref:chiba11,ref:mirollo12}
and
the noisy identical case
\cite{ref:kuramoto84,ref:manrubia04,
  ref:kuramoto81,ref:kuramoto84a,ref:mikhailov02,ref:bertini10,ref:giacomin12}.
Collective synchronization of globally coupled oscillators has also been experimentally investigated
in electrochemical oscillator systems~\cite{ref:kiss02,ref:kiss07}
or discrete chemical oscillator systems~\cite{ref:taylor09,ref:tinsley10,ref:tinsley12,ref:nkomo13}.

Recently, macroscopic synchronization
between interacting groups of globally coupled phase oscillators exhibiting collective rhythms
has also attracted considerable attention among researchers~\cite{ref:okuda91,
  ref:montbrio04,ref:abrams08,ref:barreto08,ref:sheeba08,ref:sheeba09,ref:laing09a,ref:skardal12,ref:anderson12,ref:laing12}.
Both the appearance of
the Ott-Antonsen ansatz~\cite{ref:ott08,ref:ott09,ref:ott11}
and the further understanding of
the Watanabe-Strogatz transformation~\cite{ref:pikovsky08,ref:pikovsky11,ref:marvel09,ref:watanabe93,ref:watanabe94}
have significantly facilitated theoretical investigations on collective dynamics of globally coupled oscillators
for which the phase coupling function is sinusoidal and the governing equation is deterministic (i.e., noiseless).
The Ott-Antonsen ansatz has also been applied to systems of nonlocally coupled phase oscillators
for both heterogeneous fields~\cite{ref:laing09b,ref:laing11,ref:lee11}
and homogeneous ones~\cite{ref:bordyugov10,ref:wolfrum11,ref:omelchenko13}.

To study the macroscopic synchronization between collective rhythms,
we formulated a theory for the collective phase description of globally coupled phase oscillators
\cite{ref:kawamura07,ref:kawamura08,ref:kawamura10a,ref:kawamura10b,ref:kawamura11};
the theory enables us to describe the dynamics of a collective rhythm by a single degree of freedom (i.e., a collective phase).
A similar phase description method for traveling pulses in reaction-diffusion systems was formulated~\cite{ref:nakao12},
and the collective phase description method for fully phase-locked states in networks of coupled noiseless nonidentical oscillators
was developed~\cite{ref:kori09,ref:masuda09,ref:masuda10,ref:kori12}.
We should note that there exist several investigations related to these studies;
they involve traveling pulses~\cite{ref:jalics10,ref:kilpatrick12,ref:lober12}
and fully phase-locked states~\cite{ref:ko09,ref:toenjes09,ref:cross12,ref:cross13}.

In this paper,
we consider interacting groups of globally coupled noisy limit-cycle oscillators exhibiting collective rhythms.
We formulate a collective phase description method based on the theory
developed in Refs.~\cite{ref:kawamura07,ref:kawamura08,ref:kawamura10a,ref:kawamura10b,ref:kawamura11}.
The collective phase sensitivity and collective phase coupling functions
\cite{ref:kawamura08,ref:kawamura10a,ref:kawamura11},
which quantitatively characterize the collective rhythms,
are illustrated for the first time using three representative models of limit-cycle oscillators.
Furthermore, we demonstrate a noise-induced anti-phase synchronization phenomenon,
which is predicted by the analysis of the phase model performed in Ref.~\cite{ref:kawamura10a},
by direct numerical simulations of the three limit-cycle oscillator models.

This paper is organized as follows.
In Sec.~\ref{sec:2},
we formulate a theory for the collective phase description of globally coupled noisy limit-cycle oscillators.
In Sec.~\ref{sec:3},
we illustrate the collective phase sensitivity and collective phase coupling functions using representative models.
In Sec.~\ref{sec:4},
we demonstrate noise-induced anti-phase synchronization between collective rhythms by numerical simulations.
Concluding remarks are given in Sec.~\ref{sec:5}.

\section{Formulation of the collective phase description method} \label{sec:2}

In this section,
we formulate a collective phase description method
for globally coupled noisy limit-cycle oscillators exhibiting macroscopic rhythms.
This method is based on the theory developed
in Refs.~\cite{ref:kawamura07,ref:kawamura08,ref:kawamura10a,ref:kawamura10b,ref:kawamura11}.

\subsection{Interacting groups of globally coupled noisy limit-cycle oscillators}

We consider interacting groups of globally coupled limit-cycle oscillators described by the following equation:
\begin{equation}
  \dot{\bd{X}}_j^{(\sigma)}(t)
  = \bd{F}\left( \bd{X}_j^{(\sigma)} \right)
  + \frac{1}{N} \sum_{k=1}^N \bd{G}\left( \bd{X}_j^{(\sigma)}, \bd{X}_k^{(\sigma)} \right)
  + \bd{\xi}_j^{(\sigma)}(t)
  + \epsilon_p \bd{p}_{\sigma}(t)
  + \frac{\epsilon_g}{N} \sum_{k=1}^N \bd{G}_{\sigma\tau}\left( \bd{X}_j^{(\sigma)}, \bd{X}_k^{(\tau)} \right),
  \label{eq:limit-cycle_oscillators}
\end{equation}
for $j = 1, \cdots, N$ and $(\sigma, \tau) = (1, 2)$ or $(2, 1)$,
where $\bd{X}_j^{(\sigma)}(t) \in \mathbb{R}^d$ is
the $d$-dimensional state of the $j$-th limit-cycle oscillator at time $t$
in the $\sigma$-th group consisting of $N$ elements.
The first term on the right-hand side represents the intrinsic dynamics of the limit-cycle oscillator;
the second term, the internal coupling between the oscillators within the same group;
the third term, the noise independently given to each oscillator in each group;
the fourth term, the external forcing common to all the oscillators within the $\sigma$-th group;
and the last term, the external coupling between the oscillators belonging to different groups.
The internal and external couplings are denoted by
$\bd{G}(\bd{X}_j^{(\sigma)}, \bd{X}_k^{(\sigma)})$ and $\bd{G}_{\sigma\tau}(\bd{X}_j^{(\sigma)}, \bd{X}_k^{(\tau)})$,
respectively.
The characteristic intensity of the external coupling between different groups is given by $\epsilon_g \geq 0$.
The external forcing is denoted by $\bd{p}_{\sigma}(t)$,
whose characteristic intensity is given by $\epsilon_p \geq 0$.
The noise $\bd{\xi}_j^{(\sigma)}(t)$ is assumed to be
independent white Gaussian noise~\cite{ref:risken89,ref:gardiner97},
the statistics of which are given by
\begin{equation}
  \left\langle \bd{\xi}_j^{(\sigma)}(t) \right\rangle = \bd{0}, \qquad
  \left\langle \bd{\xi}_j^{(\sigma)}(t) \bd{\xi}_k^{(\tau)}(s) \right\rangle
  = 2 \,\hat{\bd{\rm D}}\, \delta_{jk} \delta_{\sigma\tau} \delta(t - s).
\end{equation}
The diagonal matrix $\hat{\bd{\rm D}} \in \mathbb{R}^{d \times d}$
represents the noise intensity applied to each component.
We assume that the dynamics of the isolated oscillator, i.e., $\dot{\bd{X}} = \bd{F}(\bd{X})$,
is given by the following limit-cycle solution~\cite{ref:kuramoto84}:
\begin{equation}
  \bd{X}(t) = \bd{X}_0(\phi), \qquad
  \dot{\phi}(t) = \omega,
  \label{eq:limit-cycle_solution}
\end{equation}
where $\phi$ and $\omega$ are the individual phase and natural frequency, respectively.
The limit-cycle solution $\bd{X}_0(\phi)$ possesses the following $2\pi$-periodicity:
$\bd{X}_0(\phi + 2\pi) = \bd{X}_0(\phi)$.

\subsection{Interacting groups of globally coupled noisy phase oscillators}

When the perturbations applied to the individual limit-cycle oscillators given in Eq.~(\ref{eq:limit-cycle_oscillators}),
$\bd{G}$, $\bd{\xi}_j^{(\sigma)}$, $\epsilon_p \bd{p}_{\sigma}$, and $\epsilon_g \bd{G}_{\sigma\tau}$,
are sufficiently weak,
the oscillators can be described by the individual phases given in Eq.~(\ref{eq:limit-cycle_solution}).
Under the condition of weak perturbations,
we can apply the phase reduction and averaging methods to Eq.~(\ref{eq:limit-cycle_oscillators})
for approximately deriving the following individual phase equation~\cite{ref:kuramoto84,ref:kawamura07,ref:kawamura08}:
\begin{equation}
  \dot{\phi}_j^{(\sigma)}(t)
  = \omega
  + \frac{1}{N} \sum_{k=1}^N \Gamma\left( \phi_j^{(\sigma)} - \phi_k^{(\sigma)} \right)
  + \xi_j^{(\sigma)}(t)
  + \epsilon_p \bd{Z}\left( \phi_j^{(\sigma)} \right) \cdot \bd{p}_{\sigma}(t)
  + \frac{\epsilon_g}{N} \sum_{k=1}^N \Gamma_{\sigma\tau}\left( \phi_j^{(\sigma)} - \phi_k^{(\tau)} \right),
  \label{eq:phase_oscillators}
\end{equation}
where $\phi_j^{(\sigma)}(t) \in \mathbb{S}^1$ is the phase of the $j$-th oscillator at time $t$ in the $\sigma$-th group.
The first term on the right-hand side represents the frequency of the oscillators~\footnote{
  Precisely speaking, owing to the noise,
  the frequency of the oscillators given in Eq.~(\ref{eq:phase_oscillators})
  is slightly different from the natural frequency given in Eq.~(\ref{eq:limit-cycle_solution});
  however, this point is not essential in this paper.
  The theory for stochastic phase reduction of limit-cycle oscillators
  has been developed in Refs.~\cite{ref:yoshimura08,ref:teramae09,ref:nakao10,ref:goldobin10},
  with an application to common-noise-induced synchronization~\cite{ref:teramae04,ref:goldobin05,ref:nakao07,ref:kurebayashi12}.
};
the second term, the internal phase coupling function within the same group;
the third term, the effective noise;
the fourth term, the external forcing;
and the last term, the external phase coupling function between different groups.
The noise $\xi_j^{(\sigma)}(t)$ is independent white Gaussian noise~\cite{ref:risken89,ref:gardiner97},
the statistics of which are given by
\begin{equation}
  \left\langle \xi_j^{(\sigma)}(t) \right\rangle = 0, \qquad
  \left\langle \xi_j^{(\sigma)}(t) \xi_k^{(\tau)}(s) \right\rangle
  = 2 D \delta_{jk} \delta_{\sigma\tau} \delta(t - s).
\end{equation}
Here, the individual phase sensitivity function~\cite{ref:kuramoto84},
which quantifies the phase response of an individual oscillator to weak perturbations,
is denoted by $\bd{Z}(\phi_j^{(\sigma)})$.
Using this function, the internal and external phase coupling functions are given by
\begin{equation}
  \Gamma\left( \phi_j^{(\sigma)} - \phi_k^{(\sigma)} \right)
  = \frac{1}{2\pi} \int_0^{2\pi} d\lambda \, \bd{Z}\left( \lambda + \phi_j^{(\sigma)} \right) \cdot
  \bd{G}\left( \bd{X}_0\left( \lambda + \phi_j^{(\sigma)} \right), \bd{X}_0\left( \lambda + \phi_k^{(\sigma)} \right) \right)
\end{equation}
and
\begin{equation}
  \Gamma_{\sigma\tau}\left( \phi_j^{(\sigma)} - \phi_k^{(\tau)} \right)
  = \frac{1}{2\pi} \int_0^{2\pi} d\lambda \, \bd{Z}\left( \lambda + \phi_j^{(\sigma)} \right) \cdot
  \bd{G}_{\sigma\tau}\left( \bd{X}_0\left( \lambda + \phi_j^{(\sigma)} \right), \bd{X}_0\left( \lambda + \phi_k^{(\tau)} \right) \right),
\end{equation}
respectively.
In addition, the effective noise intensity is obtained as
\begin{equation}
  D = \frac{1}{2\pi} \int_0^{2\pi} d\lambda \, \bd{Z}(\lambda) \cdot \hat{\bd{\rm D}} \bd{Z}(\lambda).
\end{equation}

\subsection{Coupled nonlinear Fokker-Planck equations}

In the continuum limit, i.e., for $N \to \infty$,
the Langevin-type phase equation~(\ref{eq:phase_oscillators})
can be transformed into the following coupled nonlinear Fokker-Planck equation
\cite{ref:kuramoto84,ref:okuda91,ref:kawamura08,ref:kawamura10a,ref:kawamura11}:
\begin{align}
  \frac{\partial}{\partial t} f^{(\sigma)}(\phi, t) =
  &- \frac{\partial}{\partial \phi}
  \left[ \left\{ \omega + \int_0^{2\pi} d\phi' \, \Gamma(\phi - \phi')
    f^{(\sigma)}(\phi', t) \right\} f^{(\sigma)}(\phi, t) \right]
  + D \frac{\partial^2}{\partial \phi^2} f^{(\sigma)}(\phi, t)
  \nonumber \\
  &- \epsilon_p \frac{\partial}{\partial \phi}
  \left[ \bd{Z}(\phi) f^{(\sigma)}(\phi, t) \right] \cdot \bd{p}_{\sigma}(t)
  - \epsilon_g \frac{\partial}{\partial \phi}
  \left[ \int_0^{2\pi} d\phi' \, \Gamma_{\sigma\tau}(\phi - \phi')
    f^{(\tau)}(\phi', t) f^{(\sigma)}(\phi, t) \right].
  \label{eq:nonlinearFP}
\end{align}
Here, $f^{(\sigma)}(\phi, t)$ is
the time-dependent one-body probability density function of the phase $\phi$ in the $\sigma$-th group,
and the function is normalized as
\begin{equation}
  \int_0^{2\pi} d\phi \, f^{(\sigma)}(\phi, t) = 1.
\end{equation}
The first two terms on the right-hand side of Eq.~(\ref{eq:nonlinearFP}) represent the internal dynamics of the $\sigma$-th group,
the third term represents the external forcing applied to the $\sigma$-th group,
and the last term represents the external coupling between the $\sigma$-th group and the $\tau$-th group.
When external forcing and external coupling are absent,
each group obeying Eq.~(\ref{eq:nonlinearFP}) with $\epsilon_p = \epsilon_g = 0$
is assumed to exhibit the following collectively oscillating solution
\cite{ref:kawamura07,ref:kawamura08,ref:kawamura10a}:
\begin{equation}
  f(\phi, t) = f_0(\varphi), \qquad
  \varphi = \phi - \Theta, \qquad
  \dot{\Theta}(t) = \Omega,
\end{equation}
where $\Theta$ and $\Omega$ are the collective phase and collective frequency, respectively.
It should be noted that
the internal phase coupling function $\Gamma(\phi)$ must satisfy the in-phase condition $\Gamma'(0) < 0$,
and there is a critical noise intensity,
below which this collectively oscillating solution exists~\cite{ref:kuramoto84}.

\subsection{Coupled collective phase oscillators}

Now, let us consider external forcing and external coupling
with characteristic intensities that are sufficiently weak compared to internal coupling.
We can thus consider the third and last terms in Eq.~(\ref{eq:nonlinearFP}) as perturbations.
Under this additional condition of weak perturbations,
we can apply the phase reduction method again to Eq.~(\ref{eq:nonlinearFP})
for approximately deriving the following collective phase equation~\cite{ref:kawamura08,ref:kawamura10a,ref:kawamura11}:
\begin{equation}
  \dot{\Theta}^{(\sigma)}(t)
  = \Omega
  + \epsilon_p \bd{\zeta}\left( \Theta^{(\sigma)} \right) \cdot \bd{p}_{\sigma}(t)
  + \epsilon_g \gamma_{\sigma\tau}\left( \Theta^{(\sigma)} - \Theta^{(\tau)} \right),
\end{equation}
where $\Theta^{(\sigma)}(t) \in \mathbb{S}^1$ is the collective phase of the $\sigma$-th group at time $t$.
The first term on the right-hand side represents the collective frequency,
the second term represents the external forcing,
and the last term represents the external coupling between different groups.
The {\it collective phase sensitivity function} is given by
\begin{equation}
  \bd{\zeta}\left( \Theta^{(\sigma)} \right)
  = \int_0^{2\pi} d\varphi \, \bd{Z}\left( \varphi + \Theta^{(\sigma)} \right) k_0(\varphi),
  \label{eq:zeta}
\end{equation}
and the {\it collective phase coupling function} is given by
\begin{equation}
  \gamma_{\sigma\tau}\left( \Theta^{(\sigma)} - \Theta^{(\tau)} \right)
  = \int_0^{2\pi} d\varphi \int_0^{2\pi} d\varphi' \,
  \Gamma_{\sigma\tau}\left( \varphi - \varphi' + \Theta^{(\sigma)} - \Theta^{(\tau)} \right)
  k_0(\varphi) f_0(\varphi').
  \label{eq:gamma}
\end{equation}
Here, we defined the kernel function as
\begin{equation}
  k_0(\varphi) = -f_0(\varphi) \frac{d}{d\varphi} u_0^\ast(\varphi),
\end{equation}
which is normalized as
\begin{equation}
  \int_0^{2\pi} d\varphi \, k_0(\varphi)
  = \int_0^{2\pi} d\varphi \, u_0^\ast(\varphi) u_0(\varphi)
  = 1,
\end{equation}
where the left and right eigenfunctions associated with the zero eigenvalue are denoted by
$u_0^\ast(\varphi)$ and $u_0(\varphi) = df_0(\varphi)/d\varphi$,
respectively~\cite{ref:kawamura07,ref:kawamura08,ref:kawamura10a}.
Finally, the hierarchy of equations for the collective phase description
is summarized in Table~\ref{table:1}.
\begin{table*}[t]
  \centering
  \caption{Hierarchy of equations for the collective phase description.}
  \includegraphics[width=0.7\hsize,clip]{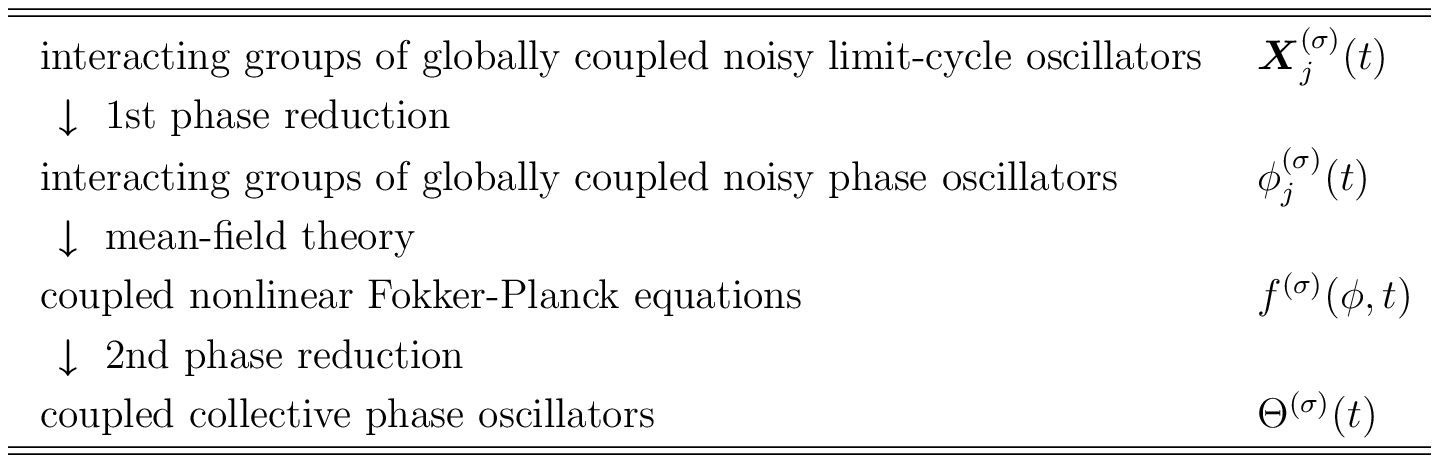}
  \label{table:1}
\end{table*}

\section{Illustration of collective phase sensitivity and coupling functions} \label{sec:3}

In this section,
we illustrate the collective phase sensitivity and coupling functions
using two representative models of limit-cycle oscillators,
Stuart-Landau oscillators and FitzHugh-Nagumo oscillators,
the state spaces of which are two-dimensional, i.e., $d = 2$.
We consider only the case in which the external coupling function is the same as the internal one.
The vector functions given in Eq.~(\ref{eq:limit-cycle_oscillators}) are thus denoted by
\begin{equation}
  \bd{X} = \begin{pmatrix} X \\ Y \end{pmatrix}, \qquad
  \bd{F} = \begin{pmatrix} F_X \\ F_Y \end{pmatrix}, \qquad
  \bd{G} = \bd{G}_{\sigma\tau} = \begin{pmatrix} G_X \\ G_Y \end{pmatrix}.
\end{equation}
The individual phase sensitivity and collective phase sensitivity functions are also denoted by
\begin{equation}
  \bd{Z} = \begin{pmatrix} Z_X \\ Z_Y \end{pmatrix}, \qquad
  \bd{\zeta} = \begin{pmatrix} \zeta_X \\ \zeta_Y \end{pmatrix}.
\end{equation}

\subsection{Stuart-Landau oscillators}

The model of Stuart-Landau oscillators is as follows.
First, the dynamics of individual oscillators is given by
\begin{subequations}
  \begin{align}
    F_X(\bd{X}) &= (X - c_0 Y) - (X^2 + Y^2) (X - c_2 Y), \\
    F_Y(\bd{X}) &= (Y + c_0 X) - (X^2 + Y^2) (Y + c_2 X).
  \end{align}
\end{subequations}
Second, the internal and external couplings are given by
\begin{subequations}
  \begin{align}
    G_X(\bd{X}, \bd{X}') &= K (X' - c_1 Y'), \\
    G_Y(\bd{X}, \bd{X}') &= K (Y' + c_1 X').
  \end{align}
\end{subequations}
Finally, the noise intensity is given by
\begin{equation}
  \hat{\bd{\rm D}} = {\rm diag}(D_0, D_0).
\end{equation}
By virtue of the circular symmetry of Stuart-Landau oscillators,
the phase reduction of Eq.~(\ref{eq:limit-cycle_oscillators}) to Eq.~(\ref{eq:phase_oscillators})
can be analytically calculated~\cite{ref:kuramoto84,ref:kawamura07,ref:kawamura08,ref:kawamura10a,ref:kawamura10b}.
The limit-cycle solution of individual Stuart-Landau oscillators is given by
\begin{subequations}
  \begin{align}
    X_0(\phi) &= \cos\phi, \\
    Y_0(\phi) &= \sin\phi,
  \end{align}
\end{subequations}
where the natural frequency is $\omega = c_0 - c_2$.
The individual phase sensitivity function is obtained as
\begin{subequations}
  \begin{align}
    Z_X(\phi) &= - \sin\phi - c_2 \cos\phi, \\
    Z_Y(\phi) &= + \cos\phi - c_2 \sin\phi.
  \end{align}
\end{subequations}
The internal and external phase coupling functions are thus given by
\begin{equation}
  \Gamma(\phi) = \Gamma_{\sigma\tau}(\phi) = -\rho \sin(\phi + \alpha), \qquad
  \rho \exp(i \alpha) = K (1 + i c_2) (1 - i c_1),
  \label{eq:sinusoidal}
\end{equation}
where the phase coupling function must satisfy with the in-phase condition $\Gamma'(0) < 0$;
that is, the phase shift must satisfy the condition $|\alpha| < \pi/2$.
In addition, the effective noise intensity is obtained as
\begin{equation}
  D = D_0 (1 + c_2^2).
\end{equation}
The parameters of the Stuart-Landau oscillators are fixed at
$c_0 = 3$, $c_1 = 0$, $c_2 = 2$, and $K = 0.05$.
Then, the critical noise intensity of $D_0$,
below which the collectively oscillating solution exists,
is exactly $0.0050$.

\begin{figure*}
  \begin{center}
    \includegraphics[width=0.9\hsize,clip]{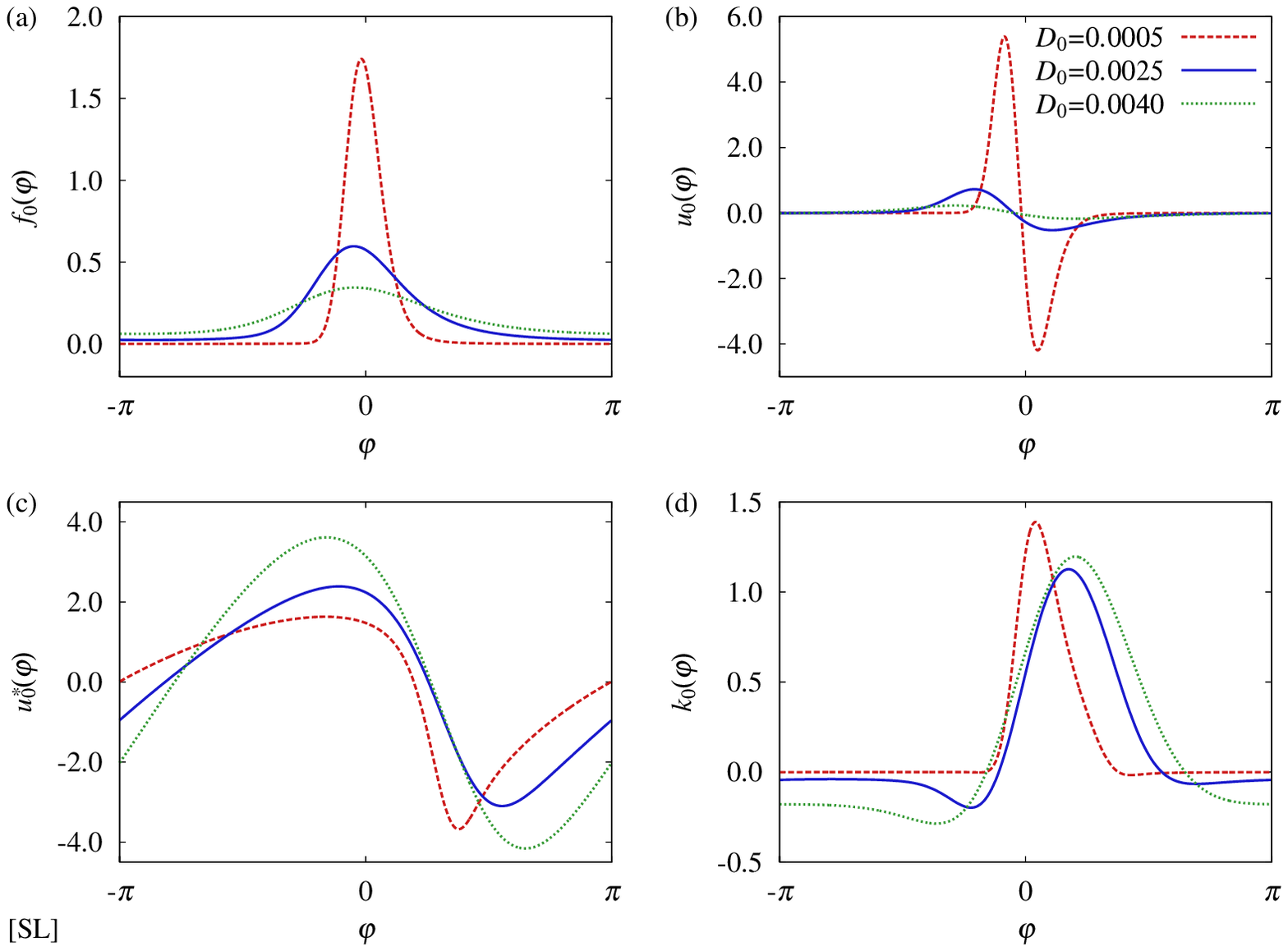}
    \caption{(Color online)
      Globally coupled Stuart-Landau (SL) oscillators.
      Parameters are $c_0 = 3$, $c_1 = 0$, $c_2 = 2$, and $K = 0.05$,
      where the critical noise intensity of $D_0$ is exactly $0.0050$.
      (a) Collectively oscillating solution $f_0(\varphi)$.
      (b) Right zero eigenfunction $u_0(\varphi)$.
      (c) Left zero eigenfunction $u_0^\ast(\varphi)$.
      (d) Kernel function $k_0(\varphi)$.
    }
    \label{fig:1}
  \end{center}
\end{figure*}

Figure~\ref{fig:1} shows
typical shapes of the collectively oscillating solution and other associated functions,
$f_0(\varphi)$, $u_0(\varphi)$, $u_0^\ast(\varphi)$, and $k_0(\varphi)$,
which were numerically obtained from the nonlinear Fokker-Planck equation~(\ref{eq:nonlinearFP})
with $\epsilon_p = \epsilon_g = 0$ and the linearized and adjoint equations.
Details of the numerical method are described in Ref.~\cite{ref:kawamura07}.

\begin{figure*}
  \begin{center}
    \includegraphics[width=0.9\hsize,clip]{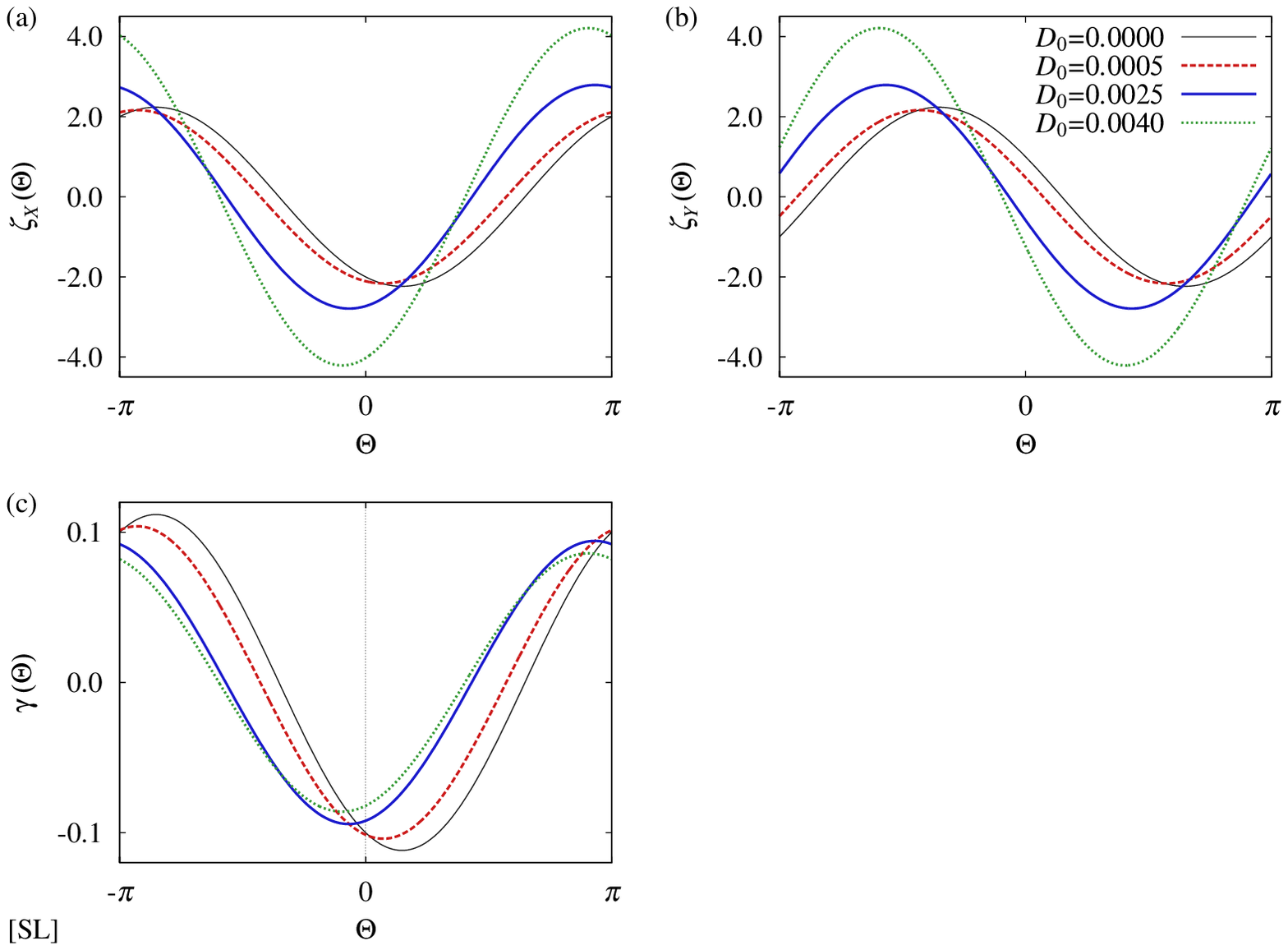}
    \caption{(Color online)
      Globally coupled Stuart-Landau (SL) oscillators.
      (a) Collective phase sensitivity function $\zeta_X(\Theta)$.
      (b) Collective phase sensitivity function $\zeta_Y(\Theta)$.
      (c) Collective phase coupling function $\gamma(\Theta)$.
    }
    \label{fig:2}
  \end{center}
\end{figure*}

The collective phase sensitivity and coupling functions,
$\bd{\zeta}(\Theta)$ and $\gamma_{\sigma\tau}(\Theta)$, are shown in Fig.~\ref{fig:2}.
For comparison,
the individual phase sensitivity and coupling functions are also shown as $D_0 = 0$.
Owing to the convolutional form given in Eq.~(\ref{eq:zeta})
and the sinusoidal form of the individual phase sensitivity function $\bd{Z}(\phi)$,
the collective phase sensitivity function $\bd{\zeta}(\Theta)$ also takes a sinusoidal form.
In the same manner,
owing to the double convolutional form given in Eq.~(\ref{eq:gamma})
and the sinusoidal form of the microscopic external phase coupling function $\Gamma_{\sigma\tau}(\phi)$,
the collective phase coupling function $\gamma_{\sigma\tau}(\Theta)$ also takes a sinusoidal form.

As seen in Fig.~\ref{fig:2}(c),
the type of the collective phase coupling function is remarkable.
The type of the microscopic external phase coupling function is in-phase coupling,
i.e., $\Gamma_{\sigma\tau}'(0) < 0$ and $\Gamma_{\sigma\tau}'(\pm\pi) > 0$.
As the noise intensity is increased,
the type of the collective phase coupling function becomes anti-phase coupling,
i.e., $\gamma_{\sigma\tau}'(0) > 0$ and $\gamma_{\sigma\tau}'(\pm\pi) < 0$.
For the sinusoidal in-phase coupling case given in Eq.~(\ref{eq:sinusoidal}),
this noise-induced anti-phase coupling can occur under the condition $\tan^2\alpha > 2$.
Details of this fact, derived from the analysis of such a phase model, are given in Ref.~\cite{ref:kawamura10a}.

\subsection{FitzHugh-Nagumo oscillators}

To confirm the general applicability of the theory,
let us consider the model of FitzHugh-Nagumo oscillators described by the following.
First, the dynamics of individual oscillators is given by
\begin{subequations}
  \begin{align}
    F_X(\bd{X}) &= \varepsilon^{-1} (X - X^3 - Y), \\
    F_Y(\bd{X}) &= a X + b.
  \end{align}
\end{subequations}
Second, the internal and external couplings are assumed to be given by
\begin{subequations}
  \begin{align}
    G_X(\bd{X}, \bd{X}') &= K_X X', \\
    G_Y(\bd{X}, \bd{X}') &= K_Y X'.
  \end{align}
\end{subequations}
Finally, the noise intensity is given by
\begin{equation}
  \hat{\bd{\rm D}} = {\rm diag}(D_0, D_0).
\end{equation}
The parameters of the FitzHugh-Nagumo oscillators are fixed at
$\varepsilon = 0.5$, $a = 1.0$, $b = 0.3$, $K_X = 0.03$, and $K_Y = 0.05$.
Then, the critical noise intensity of $D_0$,
below which the collectively oscillating solution exists,
is approximately $0.0030$.

\begin{figure*}
  \begin{center}
    \includegraphics[width=0.9\hsize,clip]{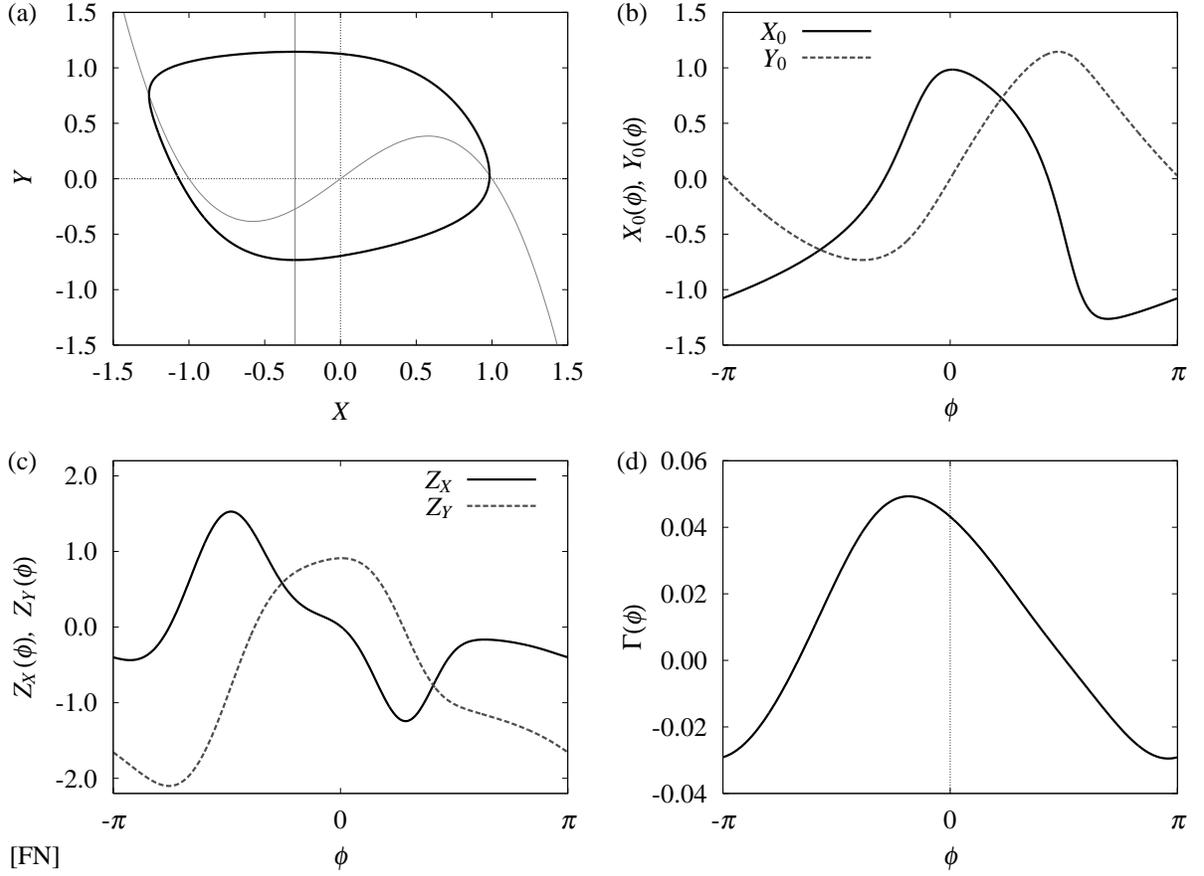}
    \caption{
      FitzHugh-Nagumo (FN) oscillators.
      Oscillator parameters are
      $\varepsilon = 0.5$, $a = 1.0$, and $b = 0.3$,
      where the natural frequency is $\omega \simeq 1.2$
      and the amplitude Floquet exponent is approximately $-2.2$.
      The coupling parameters are $K_X = 0.03$ and $K_Y = 0.05$.
      (a) Limit-cycle orbit and nullclines.
      (b) Limit-cycle solution $\bd{X}_0(\phi)$.
      (c) Phase sensitivity function $\bd{Z}(\phi)$.
      (d) Phase coupling function $\Gamma(\phi)$.
    }
    \label{fig:3}
  \end{center}
\end{figure*}

\begin{figure*}
  \begin{center}
    \includegraphics[width=0.9\hsize,clip]{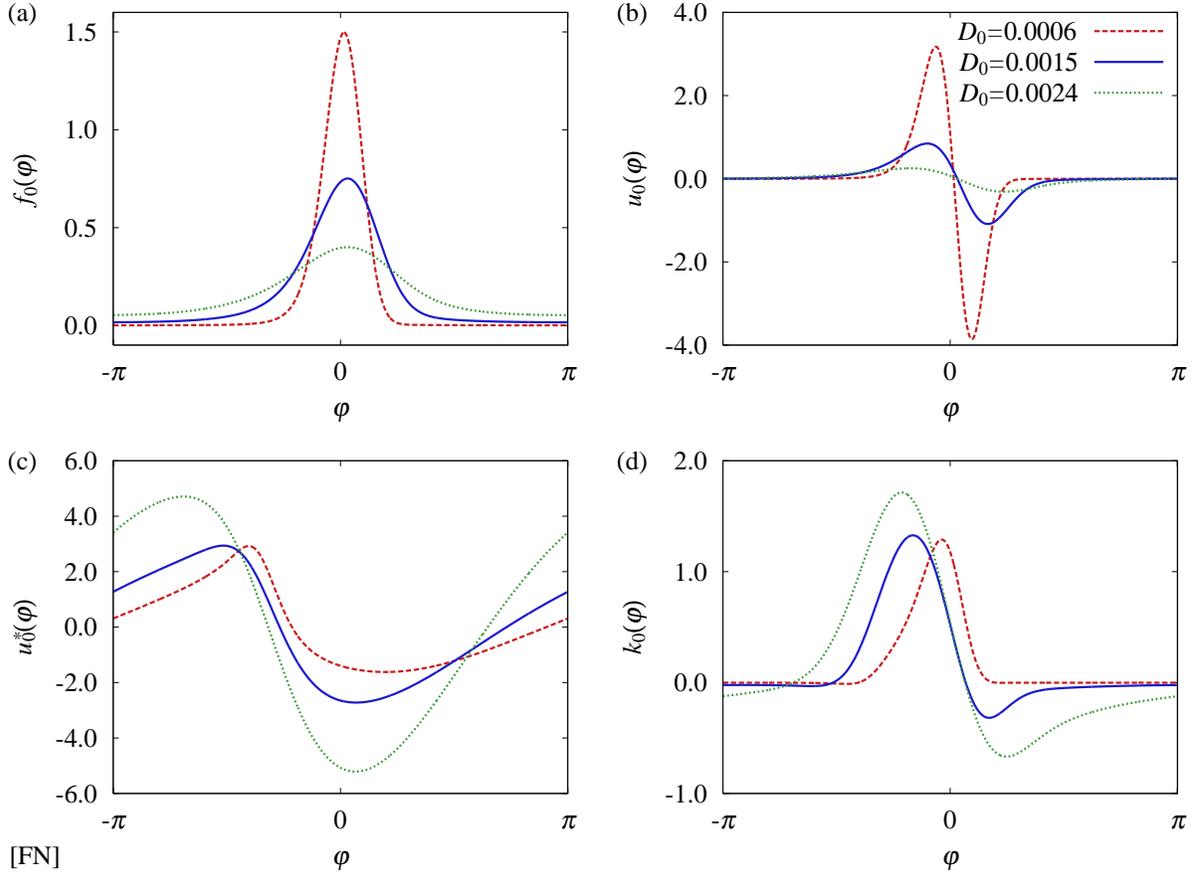}
    \caption{(Color online)
      Globally coupled FitzHugh-Nagumo (FN) oscillators.
      The critical noise intensity of $D_0$
      is approximately $0.0030$.
      (a) Collectively oscillating solution $f_0(\varphi)$.
      (b) Right zero eigenfunction $u_0(\varphi)$.
      (c) Left zero eigenfunction $u_0^\ast(\varphi)$.
      (d) Kernel function $k_0(\varphi)$.
    }
    \label{fig:4}
  \end{center}
\end{figure*}

Figure~\ref{fig:3} shows
the limit-cycle solution $\bd{X}_0(\phi)$,
the phase sensitivity function $\bd{Z}(\phi)$,
and the phase coupling function $\Gamma(\phi)$,
all of which were numerically obtained.
In contrast to the case of Stuart-Landau oscillators,
the phase sensitivity and coupling functions include higher harmonics.
Typical shapes of the collectively oscillating solution and other associated functions,
$f_0(\varphi)$, $u_0(\varphi)$, $u_0^\ast(\varphi)$, and $k_0(\varphi)$,
are shown in Fig.~\ref{fig:4};
these were obtained in a manner similar to the calculations of Fig.~\ref{fig:1}.

\begin{figure*}
  \begin{center}
    \includegraphics[width=0.9\hsize,clip]{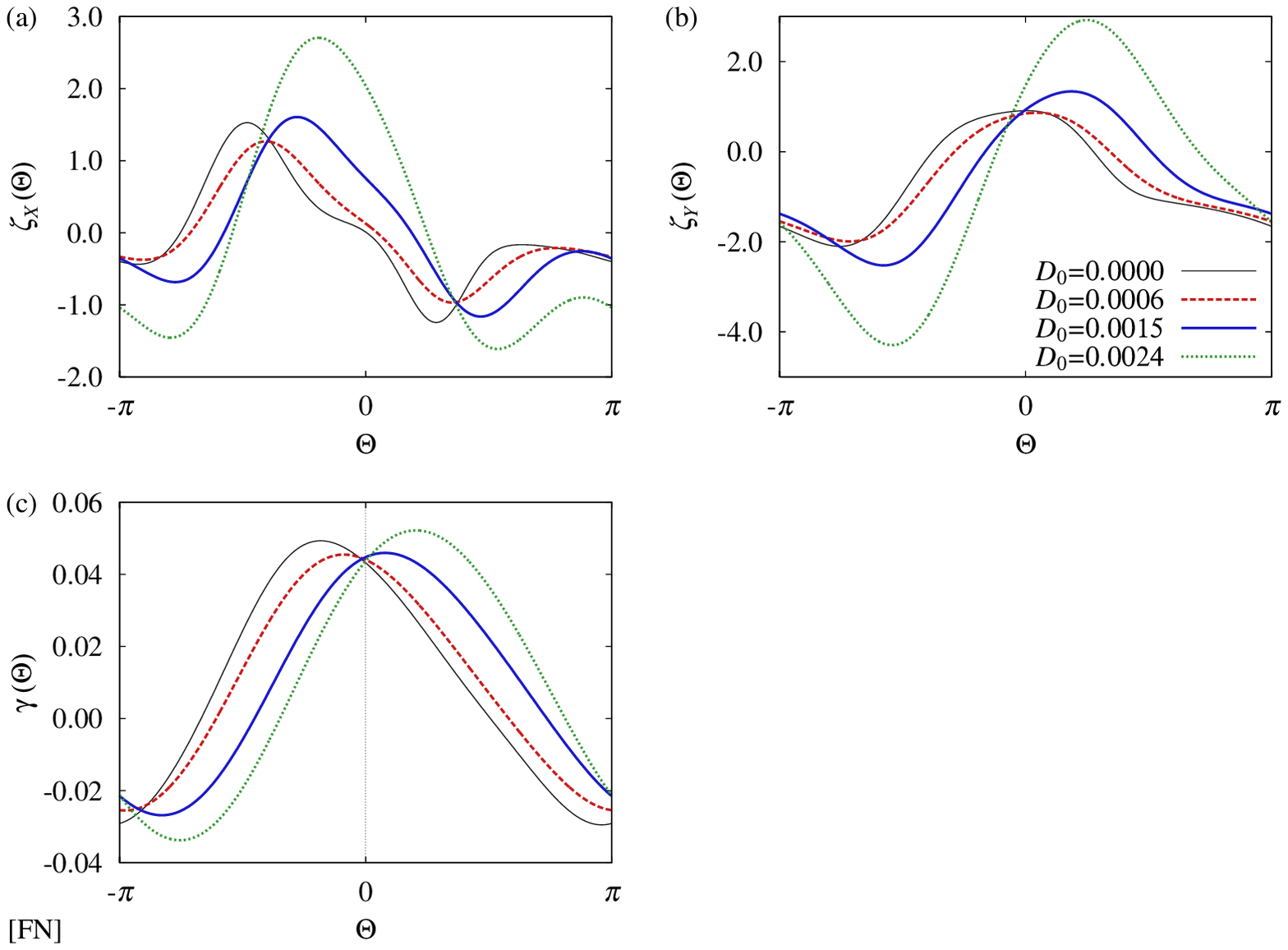}
    \caption{(Color online)
      Globally coupled FitzHugh-Nagumo (FN) oscillators.
      (a) Collective phase sensitivity function $\zeta_X(\Theta)$.
      (b) Collective phase sensitivity function $\zeta_Y(\Theta)$.
      (c) Collective phase coupling function $\gamma(\Theta)$.
    }
    \label{fig:5}
  \end{center}
\end{figure*}

The collective phase sensitivity and coupling functions,
$\bd{\zeta}(\Theta)$ and $\gamma_{\sigma\tau}(\Theta)$, are shown in Fig.~\ref{fig:5}.
In general, the collective phase sensitivity function $\bd{\zeta}(\Theta)$
possesses a smoother shape than the individual one $\bd{Z}(\phi)$
because $\bd{\zeta}(\Theta)$ is given by the convolution of $\bd{Z}(\phi)$,
as given by Eq.~(\ref{eq:zeta}).
As in the case of Stuart-Landau oscillators,
a change in the type of the collective phase coupling function can occur.
In fact, as the noise intensity is increased,
the type of the collective phase coupling function becomes anti-phase coupling,
i.e., $\gamma_{\sigma\tau}'(0) > 0$ and $\gamma_{\sigma\tau}'(\pm\pi) < 0$,
whereas the type of the microscopic external phase coupling function is in-phase coupling,
i.e., $\Gamma_{\sigma\tau}'(0) < 0$ and $\Gamma_{\sigma\tau}'(\pm\pi) > 0$.
In common with the two representative oscillator models studied above,
and presumably in more general contexts,
this transition to anti-phase coupling induced by noise
is expected to occur under the condition that
the microscopic external phase coupling function,
which is of the in-phase type,
can easily change type by a small change of parameter values.

In the following section,
we demonstrate noise-induced anti-phase synchronization between macroscopic rhythms
by direct numerical simulations of the above two models.
A realistic model for electrochemical oscillators is analyzed in Appendix~\ref{sec:A}.

\section{Demonstration of noise-induced anti-phase synchronization} \label{sec:4}

We first mention a relation between the individual phases and the collective phase.
The relation is described by the following equation for a complex order parameter:
\begin{equation}
  A^{(\sigma)}(t)
  = \frac{1}{N} \sum_{j=1}^N e^{i \phi_j^{(\sigma)}(t)}
  \simeq \int_0^{2\pi} d\phi \, e^{i \phi} f^{(\sigma)}(\phi, t)
  \simeq \int_0^{2\pi} d\phi \, e^{i \phi} f_0\left( \phi - \Theta^{(\sigma)}(t) \right)
  = R_0 e^{i \Theta^{(\sigma)}(t)},
\end{equation}
where the origin of the collective phase, i.e., $\Theta = 0$,
is determined to satisfy the following property:
\begin{equation}
  R_0 = \int_0^{2\pi} d\varphi \, e^{i \varphi} f_0(\varphi) \in \mathbb{R}.
\end{equation}
In Langevin-type simulations,
the collective phase is obtained from the individual phases
as the argument of the complex order parameter, i.e.,
$\Theta^{(\sigma)}(t) \simeq \arg A^{(\sigma)}(t)$.
In addition, for numerical simulations of limit-cycle oscillators,
the individual phase $\phi$ is calculated from the state vector $\bd{X}$
using the asymptotic phase field $\phi(\bd{X})$ prepared in advance.
Details of the numerical method are described in Ref.~\cite{ref:kawamura07}.

Here, we focus on the case in which the external forcing is absent, i.e., $\epsilon_p = 0$.
Considering two weakly interacting groups of globally coupled noisy limit-cycle oscillators,
we performed numerical simulations of Eq.~(\ref{eq:limit-cycle_oscillators})
with $\epsilon_p = 0$ and $\epsilon_g = 0.02$.
The number of oscillators in each group was $N = 10000$,
which was sufficiently large to observe clear collective oscillations.
We separately prepared two groups of limit-cycle oscillators exhibiting collective oscillations
and used these states as the initial conditions of the simulations.

Figure~\ref{fig:6} shows
the time evolution of the collective phase difference $| \Theta^{(1)} - \Theta^{(2)} |$
from an almost in-phase synchronized state of the groups.
As expected from both Fig.~\ref{fig:2}(c) for Stuart-Landau oscillators with $D_0 = 0.0025$
and Fig.~\ref{fig:5}(c) for FitzHugh-Nagumo oscillators with $D_0 = 0.0015$,
the collective phase difference $| \Theta^{(1)} - \Theta^{(2)} |$ eventually approached $\pi$;
that is, the two groups exhibited anti-phase synchronization after some time.

Snapshots of the individual limit-cycle oscillators
taken after the collective phase difference had reached the asymptotic value
are shown in Fig.~\ref{fig:7}.
The two distributions of the limit-cycle oscillators are clearly separated;
this indicates anti-phase synchronization between the groups.

\section{Concluding remarks} \label{sec:5}

In this paper,
we considered interacting groups of globally coupled noisy limit-cycle oscillators,
and we formulated the collective phase description method in a systematic manner.
The collective phase sensitivity and coupling functions
were illustrated using three representative models of limit-cycle oscillators,
i.e., Stuart-Landau oscillators, FitzHugh-Nagumo oscillators, and electrochemical oscillators.
In particular,
we demonstrated noise-induced anti-phase synchronization between macroscopic rhythms
by direct numerical simulations of the three models.
We hope that this noise-induced transition phenomenon
will be experimentally confirmed in the near future~\footnote{
  In this footnote,
  we remark the dependence of the synchronization between interacting groups
  on the external coupling intensity $\epsilon_g$.
  We focus on the case in which the external coupling function is the same as the internal one.
  When the characteristic intensity of the external coupling is also the same as that of the internal coupling,
  i.e., $\epsilon_g = 1$,
  the internal and external couplings are indistinguishable,
  and therefore,
  the groups exhibit in-phase collective synchronization.
  Thus, for the case in which the type of the collective phase coupling function is anti-phase coupling,
  the synchronization between the groups changes from in-phase to anti-phase
  as the external coupling intensity $\epsilon_g$ decreases from one.
  This transition depending on the external coupling intensity $\epsilon_g$
  cannot be analyzed using the collective phase description method
  because the amplitude degrees of freedom of collective oscillations are essential for strongly interacting cases.
  It is an interesting open problem to determine the critical value of the external coupling intensity.
}.

Finally, we emphasize that
the type of the collective phase coupling function can be different from
that of the microscopic external phase coupling function.
This fact indicates that macroscopic synchronization properties cannot be fully clarified
by simply analyzing the external coupling between individual oscillators belonging to different groups.
The collective phase description method is thus necessary and powerful for theoretical and experimental analysis
of the synchronization properties of collective rhythms arising from interacting individual oscillators.

\begin{acknowledgments}
  The author is grateful to Yoshiki Kuramoto, Hiroya Nakao, Hiroshi Kori, and Kensuke Arai for fruitful discussions.
  This work was supported by JSPS KAKENHI Grant Number 25800222.
\end{acknowledgments}

\appendix

\section{Electrochemical oscillators} \label{sec:A}

In this appendix,
we consider a realistic model for electrochemical oscillators~\cite{ref:haim92,ref:kiss08}.
The state space of the electrochemical oscillator model is two-dimensional,
and $X$ is the dimensionless double layer potential drop
and $Y$ is the surface coverage of ${\rm NiO + NiOH}$.
Perturbations such as couplings, noise, and forcing can be applied only to the $X$-component,
so that we use only the $X$-component of individual and collective phase sensitivity functions,
$Z_X$ and $\zeta_X$.
The model of electrochemical oscillators is as follows.
First, the dynamics of individual oscillators is given by
\begin{subequations}
  \begin{align}
    F_X(\bd{X})
    &= \frac{v - X}{r} - \left( \frac{C_{\rm h} \exp(0.5 X)}{1 + C_{\rm h} \exp(X)} + a \exp(X) \right) (1 - Y), \\
    F_Y(\bd{X})
    &= \frac{1}{s} \left( \frac{\exp(0.5 X)}{1 + C_{\rm h} \exp(X)} (1 - Y)
    - \frac{b \, C_{\rm h} \exp(2 X)}{c \, C_{\rm h} + \exp(X)} Y \right).
  \end{align}
\end{subequations}
Second, the internal and external couplings are given by
\begin{subequations}
  \begin{align}
    G_X(\bd{X}, \bd{X}') &= \frac{K}{r} (X' - X), \\
    G_Y(\bd{X}, \bd{X}') &= 0.
  \end{align}
\end{subequations}
Finally, the noise intensity is given by
\begin{equation}
  \hat{\bd{\rm D}} = {\rm diag}(D_0, 0).
\end{equation}
As in Ref.~\cite{ref:kiss08},
the parameters of the electrochemical oscillators are fixed at
$a = 0.3$, $b = 0.00006$, $c = 0.001$, $C_{\rm h} = 1600$,
$s = 0.01$, $r = 20.0$, $v = 15.0$, and $K/r = 0.0025$.

\begin{figure*}
  \begin{center}
    \includegraphics[width=0.9\hsize,clip]{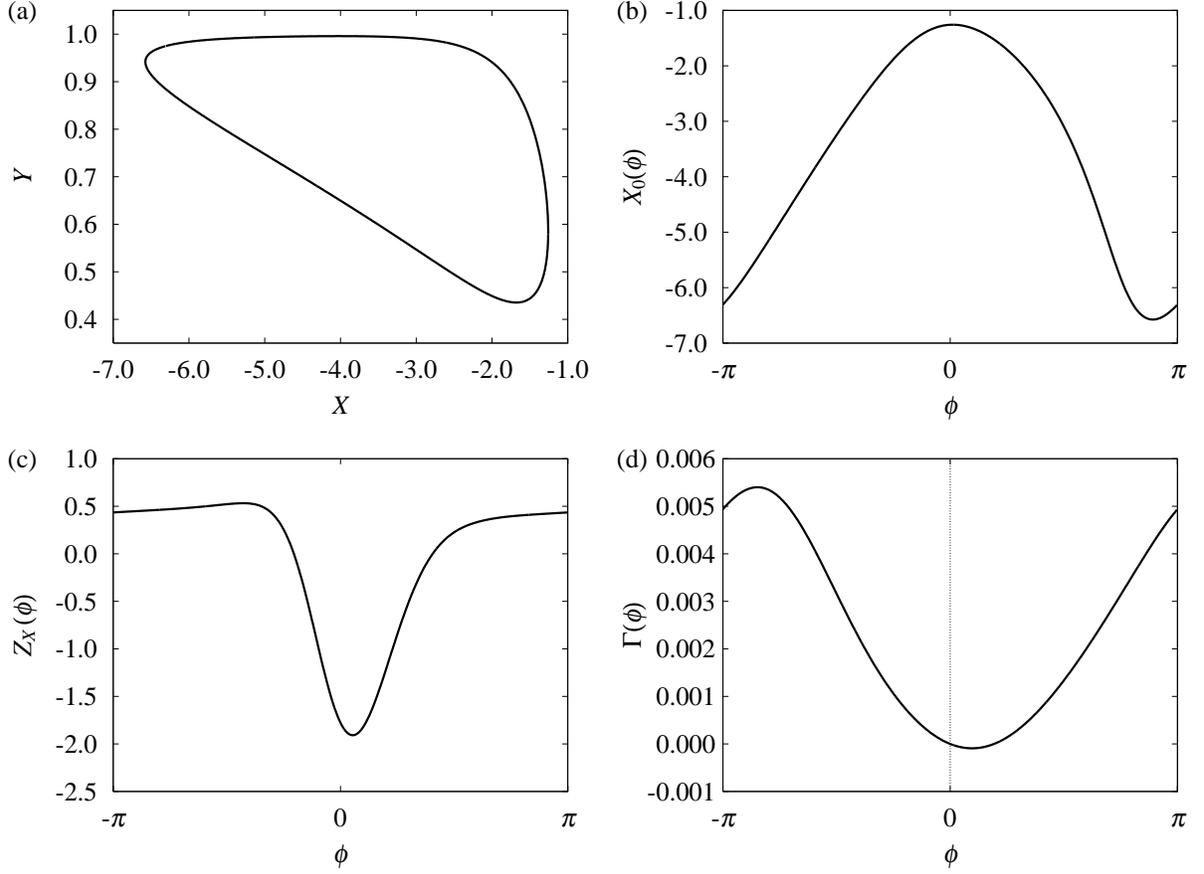}
    \caption{
      Electrochemical oscillators.
      Oscillator parameters are
      $a = 0.3$, $b = 0.00006$, $c = 0.001$, $C_{\rm h} = 1600$,
      $s = 0.01$, $r = 20.0$, and $v = 15.0$,
      where the natural frequency is $\omega \simeq 0.45$
      and the amplitude Floquet exponent is approximately $-0.23$.
      The coupling parameter is $K/r = 0.0025$.
      (a) Limit-cycle orbit.
      (b) Limit-cycle solution $X_0(\phi)$.
      (c) Phase sensitivity function $Z_X(\phi)$.
      (d) Phase coupling function $\Gamma(\phi)$.
    }
    \label{fig:A1}
  \end{center}
\end{figure*}

\begin{figure*}
  \begin{center}
    \includegraphics[width=0.9\hsize,clip]{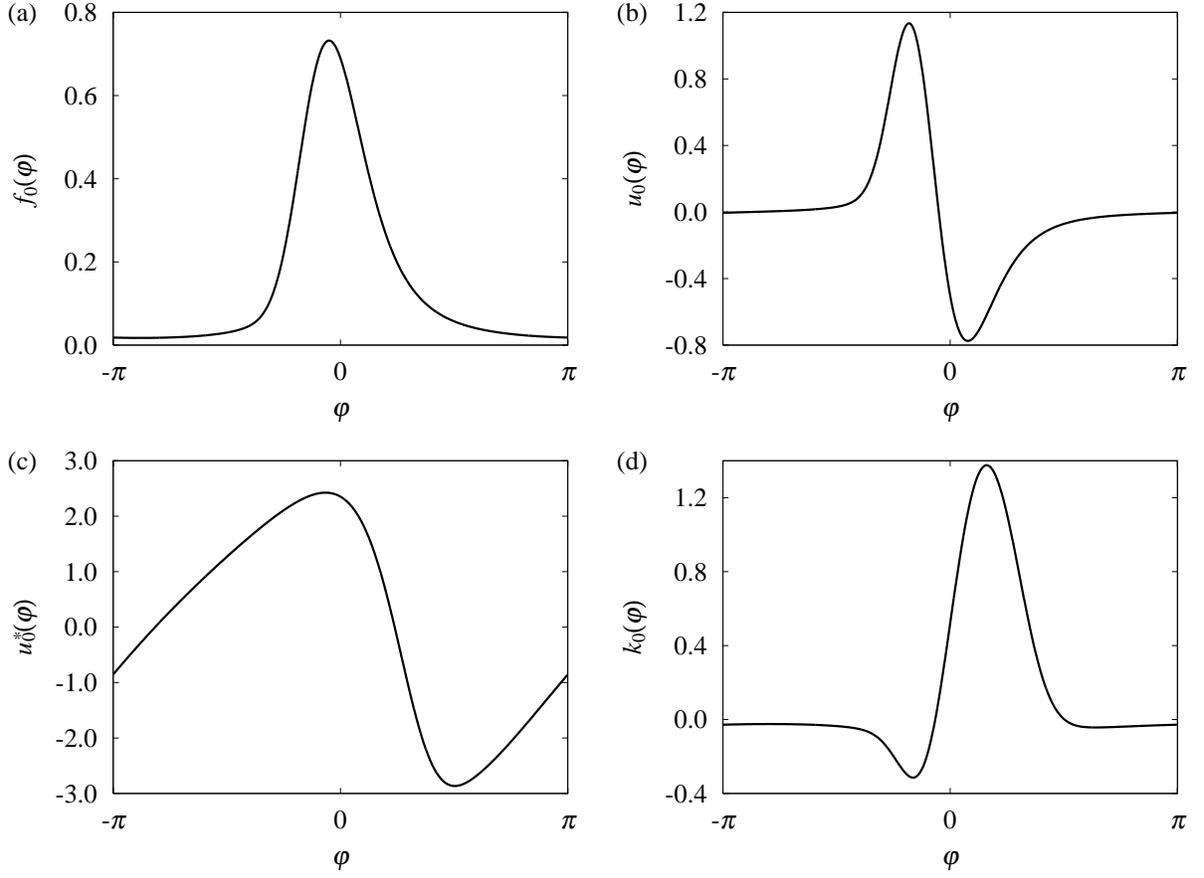}
    \caption{
      Globally coupled electrochemical oscillators.
      The noise intensity is $D/D_{\rm c} = 0.3$,
      i.e., $D_0 \simeq 0.00025$,
      where $D_{\rm c}$ is the critical value
      of the effective noise intensity $D$.
      (a) Collectively oscillating solution $f_0(\varphi)$.
      (b) Right zero eigenfunction $u_0(\varphi)$.
      (c) Left zero eigenfunction $u_0^\ast(\varphi)$.
      (d) Kernel function $k_0(\varphi)$.
    }
    \label{fig:A2}
  \end{center}
\end{figure*}

Figure~\ref{fig:A1} shows
the limit-cycle solution $X_0(\phi)$,
the phase sensitivity function $Z_X(\phi)$,
and the phase coupling function $\Gamma(\phi)$.
Under the condition $D_0 \simeq 0.00025$,
the shapes of the collectively oscillating solution and other associated functions,
$f_0(\varphi)$, $u_0(\varphi)$, $u_0^\ast(\varphi)$, and $k_0(\varphi)$,
are shown in Fig.~\ref{fig:A2}.

\begin{figure*}
  \begin{center}
    \includegraphics[width=0.9\hsize,clip]{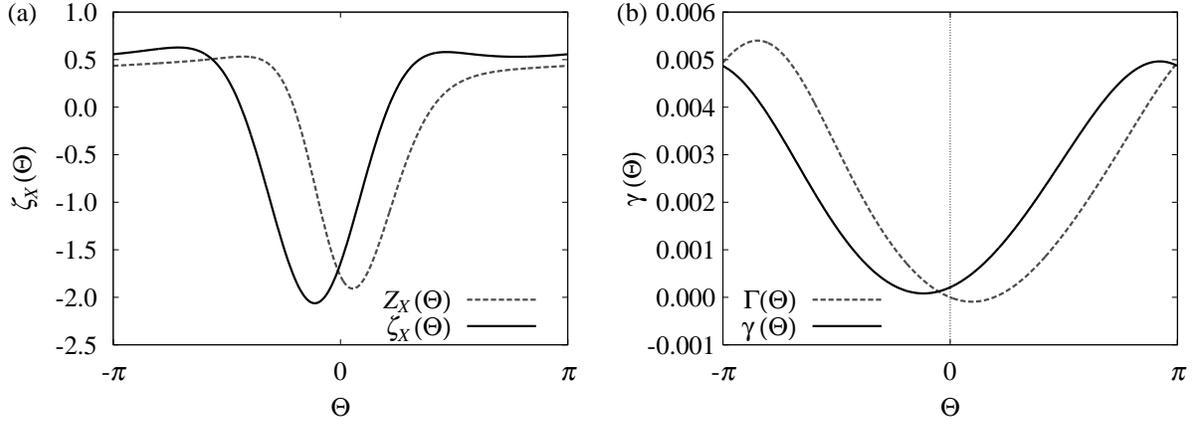}
    \caption{
      Globally coupled electrochemical oscillators.
      (a) Collective phase sensitivity function $\zeta_X(\Theta)$.
      (b) Collective phase coupling function $\gamma(\Theta)$.
    }
    \label{fig:A3}
  \end{center}
\end{figure*}

Figure~\ref{fig:A3} shows
the collective phase sensitivity and coupling functions,
$\zeta_X(\Theta)$ and $\gamma_{\sigma\tau}(\Theta)$.
The type of the collective phase coupling function is anti-phase coupling,
i.e., $\gamma_{\sigma\tau}'(0) > 0$ and $\gamma_{\sigma\tau}'(\pm\pi) < 0$,
whereas the type of the microscopic external phase coupling function is in-phase coupling,
i.e., $\Gamma_{\sigma\tau}'(0) < 0$ and $\Gamma_{\sigma\tau}'(\pm\pi) > 0$.

\begin{figure*}
  \begin{center}
    \includegraphics[width=0.9\hsize,clip]{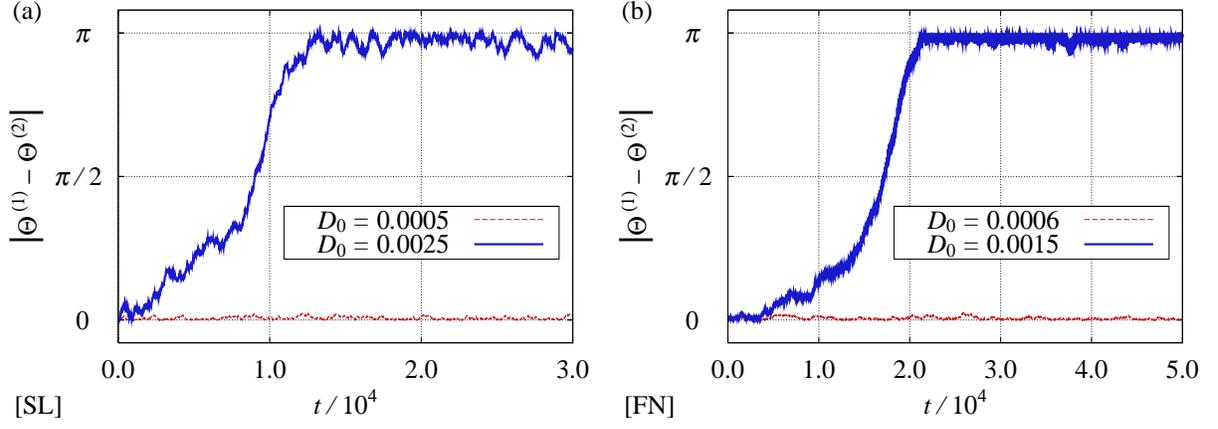}
    \caption{(Color online)
      Time evolution of the collective phase difference
      $| \Theta^{(1)} - \Theta^{(2)} |$,
      which demonstrates noise-induced anti-phase synchronization
      between collective oscillations.
      The external coupling parameter is $\epsilon_g = 0.02$.
      The number of oscillators in each group is $N = 10000$.
      (a) Globally coupled Stuart-Landau (SL) oscillators.
      (b) Globally coupled FitzHugh-Nagumo (FN) oscillators.
    }
    \label{fig:6}
  \end{center}
\end{figure*}

\begin{figure*}
  \begin{center}
    \includegraphics[width=0.9\hsize,clip]{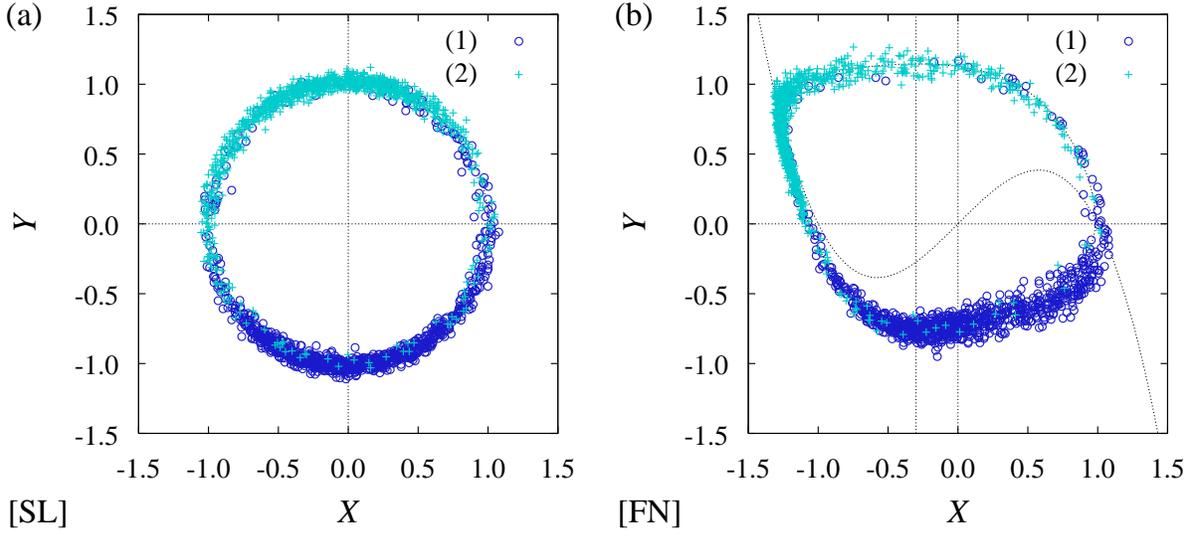}
    \caption{(Color online)
      Snapshots of individual oscillators
      exhibiting noise-induced anti-phase synchronization.
      Open circles ($\circ$) and pluses ($+$) indicate
      oscillators of group~$(1)$ and group~$(2)$, respectively.
      Only one in every ten oscillators is plotted.
      (a) Globally coupled Stuart-Landau (SL) oscillators with $D_0 = 0.0025$.
      (b) Globally coupled FitzHugh-Nagumo (FN) oscillators with $D_0 = 0.0015$.
    }
    \label{fig:7}
  \end{center}
\end{figure*}

\begin{figure*}
  \begin{center}
    \includegraphics[width=0.9\hsize,clip]{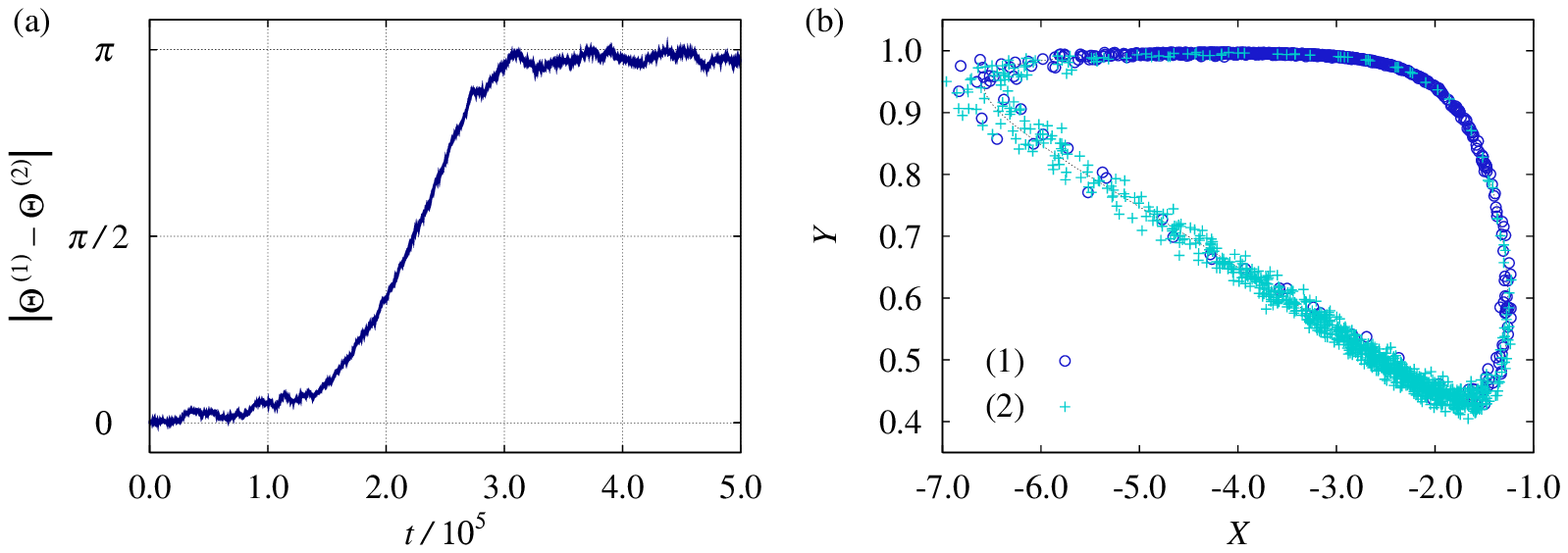}
    \caption{(Color online)
      Interacting groups of
      globally coupled electrochemical oscillators.
      The external coupling parameter is $\epsilon_g = 0.02$.
      The number of oscillators in each group is $N = 10000$.
      (a) Time evolution of the collective phase difference
      $| \Theta^{(1)} - \Theta^{(2)} |$.
      (b) Snapshot of the asymptotic state of individual oscillators.
      Open circles ($\circ$) and pluses ($+$) indicate
      oscillators of group~$(1)$ and group~$(2)$, respectively.
      Only one in every ten oscillators is plotted.
    }
    \label{fig:A4}
  \end{center}
\end{figure*}

As in Fig.~\ref{fig:6},
the time evolution of the collective phase difference $| \Theta^{(1)} - \Theta^{(2)} |$
from an almost in-phase synchronized state of the groups is shown in Fig.~\ref{fig:A4}(a),
which was obtained from a direct numerical simulation of Eq.~(\ref{eq:limit-cycle_oscillators})
with $\epsilon_p = 0$, $\epsilon_g = 0.02$, and $N = 10000$.
The collective phase difference $| \Theta^{(1)} - \Theta^{(2)} |$ eventually became $\pi$;
that is, the two groups exhibited anti-phase synchronization after some time.
As in Fig.~\ref{fig:7},
a snapshot of the individual limit-cycle oscillators
taken after the collective phase difference had reached the asymptotic value is shown in Fig.~\ref{fig:A4}(b).
The two distributions of the limit-cycle oscillators are distinctly separated;
this signifies anti-phase synchronization between the groups.

\clearpage


\clearpage


\clearpage

\clearpage

\begin{thebibliography}{99}

\bibitem{ref:winfree80}
  A.T.~Winfree,
  The Geometry of Biological Time,
  Springer, New York, 1980; Springer, Second Edition, New York, 2001.

\bibitem{ref:kuramoto84}
  Y.~Kuramoto,
  Chemical Oscillations, Waves, and Turbulence,
  Springer, New York, 1984; Dover, New York, 2003.

\bibitem{ref:pikovsky01}
  A.~Pikovsky, M.~Rosenblum, J.~Kurths,
  Synchronization: A Universal Concept in Nonlinear Sciences,
  Cambridge University Press, Cambridge, 2001.

\bibitem{ref:strogatz03}
  S.H.~Strogatz,
  Sync: How Order Emerges from Chaos in the Universe, Nature, and Daily Life,
  Hyperion Books, New York, 2003.

\bibitem{ref:manrubia04}
  S.C.~Manrubia, A.S.~Mikhailov, D.H.~Zanette,
  Emergence of Dynamical Order: Synchronization Phenomena in Complex Systems,
  World Scientific, Singapore, 2004.

\bibitem{ref:osipov07}
  G.V.~Osipov, J.~Kurths, C.~Zhou,
  Synchronization in Oscillatory Networks,
  Springer, New York, 2007.

\bibitem{ref:mikhailov13}
  A.S.~Mikhailov, G.~Ertl (Editors),
  Engineering of Chemical Complexity,
  World Scientific, Singapore, 2013.

\bibitem{ref:ermentrout96}
  G.B.~Ermentrout,
  Type I membranes, phase resetting curves, and synchrony,
  Neural Comput. 8 (1996) 979-1001.

\bibitem{ref:brown04}
  E.~Brown, J.~Moehlis, P.~Holmes,
  On the phase reduction and response dynamics of neural oscillator populations,
  Neural Comput. 16 (2004) 673-715.

\bibitem{ref:hoppensteadt97}
  F.C.~Hoppensteadt, E.M.~Izhikevich,
  Weakly Connected Neural Networks,
  Springer, New York, 1997.

\bibitem{ref:izhikevich07}
  E.M.~Izhikevich,
  Dynamical Systems in Neuroscience: The Geometry of Excitability and Bursting,
  MIT Press, Cambridge, MA, 2007.

\bibitem{ref:ermentrout10}
  G.B.~Ermentrout, D.H.~Terman,
  Mathematical Foundations of Neuroscience,
  Springer, New York, 2010.

\bibitem{ref:schultheiss12}
  N.~Schultheiss, R.~Butera, A.~Prinz (Editors),
  Phase Response Curves in Neuroscience: Theory, Experiment, and Analysis,
  Springer, New York, 2012.

\bibitem{ref:strogatz00}
  S.H.~Strogatz,
  From Kuramoto to Crawford: Exploring the onset of synchronization in populations of coupled oscillators,
  Physica D 143 (2000) 1-20.

\bibitem{ref:acebron05}
  J.A.~Acebr\'on, L.L.~Bonilla, C.J.~P\'erez Vicente, F.~Ritort, R.~Spigler,
  The Kuramoto model: A simple paradigm for synchronization phenomena,
  Rev. Mod. Phys. 77 (2005) 137-185.



\bibitem{ref:boccaletti06}
  S.~Boccaletti, V.~Latora, Y.~Moreno, M.~Chavez, D.-U.~Hwang,
  Complex networks: Structure and dynamics,
  Phys. Rep. 424 (2006) 175-308.

\bibitem{ref:arenas08}
  A.~Arenas, A.~D\'{\i}az-Guilera, J.~Kurths, Y.~Moreno, C.~Zhou,
  Synchronization in complex networks,
  Phys. Rep. 469 (2008) 93-153.


\bibitem{ref:barrat08}
  A.~Barrat, M.~Barthelemy, A.~Vespignani,
  Dynamical Processes on Complex Networks,
  Cambridge University Press, Cambridge, 2008.

\bibitem{ref:kuramoto75}
  Y.~Kuramoto,
  Self-entrainment of a population of coupled non-linear oscillators,
  in International Symposium on Mathematical Problems in Theoretical Physics, edited by H.~Araki,
  Lecture Notes in Physics Vol. 39, Springer, New York, 1975, pp. 420-422.

\bibitem{ref:sakaguchi86}
  H.~Sakaguchi, Y.~Kuramoto,
  A soluble active rotator model showing phase transitions via mutual entrainment,
  Prog. Theor. Phys. 76 (1986) 576-581.

\bibitem{ref:strogatz91}
  S.H.~Strogatz, R.E.~Mirollo,
  Stability of incoherence in a population of coupled oscillators,
  J. Stat. Phys. 63 (1991) 613-635.

\bibitem{ref:crawford95}
  J.D.~Crawford,
  Scaling and singularities in the entrainment of globally coupled oscillators,
  Phys. Rev. Lett. 74 (1995) 4341-4344.

\bibitem{ref:crawford99}
  J.D.~Crawford and K.T.R.~Davies,
  Synchronization of globally coupled phase oscillators: Singularities and scaling for general couplings,
  Physica D 125 (1999) 1-46.

\bibitem{ref:mirollo07}
  R.E.~Mirollo, S.H.~Strogatz,
  The spectrum of the partially locked state for the Kuramoto model,
  J. Nonlinear Sci. 17 (2007) 309-347.

\bibitem{ref:chiba11}
  H.~Chiba, I.~Nishikawa,
  Center manifold reduction for large populations of globally coupled phase oscillators,
  Chaos 21 (2011) 043103.

\bibitem{ref:mirollo12}
  R.E.~Mirollo,
  The asymptotic behavior of the order parameter for the infinite-$N$ Kuramoto model,
  Chaos 22 (2012) 043118.

\bibitem{ref:kuramoto81}
  Y.~Kuramoto,
  Rhythms and turbulence in populations of chemical oscillators,
  Physica A 106 (1981) 128-143.

\bibitem{ref:kuramoto84a}
  Y.~Kuramoto,
  Cooperative dynamics of oscillator community: A study based on lattice of rings,
  Prog. Theor. Phys. Suppl. 79 (1984) 223-240.

\bibitem{ref:mikhailov02}
  A.S.~Mikhailov, V.~Calenbuhr,
  From Cells to Societies: Models of Complex Coherent Action,
  Springer, New York, 2002.

\bibitem{ref:bertini10}
  L.~Bertini, G.~Giacomin, K.~Pakdaman,
  Dynamical aspects of mean field plane rotators and the Kuramoto model,
  J. Stat. Phys. 138 (2010) 270-290.

\bibitem{ref:giacomin12}
  G.~Giacomin, K.~Pakdaman, X.~Pellegrin,
  Global attractor and asymptotic dynamics in the Kuramoto model for coupled noisy phase oscillators,
  Nonlinearity 25 (2012) 1247.

\bibitem{ref:kiss02}
  I.Z.~Kiss, Y.~Zhai, J.L.~Hudson,
  Emerging coherence in a population of chemical oscillators,
  Science 296 (2002) 1676-1678.

\bibitem{ref:kiss07}
  I.Z.~Kiss, C.G.~Rusin, H.~Kori, J.L.~Hudson,
  Engineering complex dynamical structures: Sequential patterns and desynchronization,
  Science 316 (2007) 1886-1889.

\bibitem{ref:taylor09}
  A.F.~Taylor, M.R.~Tinsley, F.~Wang, Z.~Huang, K.~Showalter,
  Dynamical quorum sensing and synchronization in large populations of chemical oscillators,
  Science 323 (2009) 614-617.


\bibitem{ref:tinsley10}
  M.R.~Tinsley, A.F.~Taylor, Z.~Huang, F.~Wang, K.~Showalter,
  Dynamical quorum sensing and synchronization in collections of excitable and oscillatory catalytic particles,
  Physica D 239 (2010) 785-790.

\bibitem{ref:tinsley12}
  M.R.~Tinsley, S.~Nkomo, K.~Showalter,
  Chimera and phase-cluster states in populations of coupled chemical oscillators,
  Nature Physics 8 (2012) 662-665.

\bibitem{ref:nkomo13}
  S.~Nkomo, M.R.~Tinsley, K.~Showalter,
  Chimera states in populations of nonlocally coupled chemical oscillators,
  Phys. Rev. Lett. 110 (2013) 244102.

\bibitem{ref:okuda91}
  K.~Okuda, Y.~Kuramoto,
  Mutual entrainment between populations of coupled oscillators,
  Prog. Theor. Phys. 86 (1991) 1159-1176.

\bibitem{ref:montbrio04}
  E.~Montbri\'o, J.~Kurths, B.~Blasius,
  Synchronization of two interacting populations of oscillators,
  Phys. Rev. E 70 (2004) 056125.

\bibitem{ref:abrams08}
  D.M.~Abrams, R.E.~Mirollo, S.H.~Strogatz, D.A.~Wiley,
  Solvable model for chimera states of coupled oscillators,
  Phys. Rev. Lett. 101 (2008) 084103.

\bibitem{ref:barreto08}
  E.~Barreto, B.~Hunt, E.~Ott, P.~So,
  Synchronization in networks of networks:
  The onset of coherent collective behavior in systems of interacting populations of heterogeneous oscillators,
  Phys. Rev. E 77 (2008) 036107.

\bibitem{ref:sheeba08}
  J.H.~Sheeba, V.K.~Chandrasekar, A.~Stefanovska, P.V.E.~McClintock,
  Routes to synchrony between asymmetrically interacting oscillator ensembles,
  Phys. Rev. E 78 (2008) 025201(R).

\bibitem{ref:sheeba09}
  J.H.~Sheeba, V.K.~Chandrasekar, A.~Stefanovska, P.V.E.~McClintock,
  Asymmetry-induced effects in coupled phase-oscillator ensembles: Routes to synchronization,
  Phys. Rev. E 79 (2009) 046210.

\bibitem{ref:laing09a}
  C.R.~Laing,
  Chimera states in heterogeneous networks,
  Chaos 19 (2009) 013113.

\bibitem{ref:skardal12}
  P.S.~Skardal, J.G.~Restrepo,
  Hierarchical synchrony of phase oscillators in modular networks,
  Phys. Rev. E 85 (2012) 016208.

\bibitem{ref:anderson12}
  D.~Anderson, A.~Tenzer, G.~Barlev, M.~Girvan, T.M.~Antonsen, E.~Ott,
  Multiscale dynamics in communities of phase oscillators,
  Chaos 22 (2012) 013102.

\bibitem{ref:laing12}
  C.R.~Laing,
  Disorder-induced dynamics in a pair of coupled heterogeneous phase oscillator networks,
  Chaos 22 (2012) 043104.

\bibitem{ref:ott08}
  E.~Ott, T.M.~Antonsen,
  Low dimensional behavior of large systems of globally coupled oscillators,
  Chaos 18 (2008) 037113.

\bibitem{ref:ott09}
  E.~Ott, T.M.~Antonsen,
  Long time evolution of phase oscillator systems,
  Chaos 19 (2009) 023117.

\bibitem{ref:ott11}
  E.~Ott, B.R.~Hunt, T.M.~Antonsen,
  Comment on ``Long time evolution of phase oscillator systems'' [Chaos 19 (2009) 023117],
  Chaos 21 (2011) 025112.

\bibitem{ref:pikovsky08}
  A.~Pikovsky, M.~Rosenblum,
  Partially integrable dynamics of hierarchical populations of coupled oscillators,
  Phys. Rev. Lett. 101 (2008) 264103.

\bibitem{ref:pikovsky11}
  A.~Pikovsky, M.~Rosenblum,
  Dynamics of heterogeneous oscillator ensembles in terms of collective variables,
  Physica D 240 (2011) 872-881.

\bibitem{ref:marvel09}
  S.A.~Marvel, R.E.~Mirollo, S.H.~Strogatz,
  Identical phase oscillators with global sinusoidal coupling evolve by M\"obius group action,
  Chaos 19 (2009) 043104.

\bibitem{ref:watanabe93}
  S.~Watanabe, S.H.~Strogatz,
  Integrability of a globally coupled oscillator array,
  Phys. Rev. Lett. 70 (1993) 2391-2394.

\bibitem{ref:watanabe94}
  S.~Watanabe, S.H.~Strogatz,
  Constants of motion for superconducting Josephson arrays,
  Physica D 74 (1994) 197-253.

\bibitem{ref:laing09b}
  C.R.~Laing,
  The dynamics of chimera states in heterogeneous Kuramoto networks,
  Physica D 238 (2009) 1569-1588.

\bibitem{ref:laing11}
  C.R.~Laing,
  Fronts and bumps in spatially extended Kuramoto networks,
  Physica D 240 (2011) 1960-1971.

\bibitem{ref:lee11}
  W.S.~Lee, J.G.~Restrepo, E.~Ott, T.M.~Antonsen,
  Dynamics and pattern formation in large systems of spatially-coupled oscillators with finite response times,
  Chaos 21 (2011) 023122.

\bibitem{ref:bordyugov10}
  G.~Bordyugov, A.~Pikovsky, M.~Rosenblum,
  Self-emerging and turbulent chimeras in oscillator chains,
  Phys. Rev. E 82 (2010) 035205(R).

\bibitem{ref:wolfrum11}
  M.~Wolfrum, O.E.~Omel'chenko, S.~Yanchuk, Y.L.~Maistrenko,
  Spectral properties of chimera states,
  Chaos 21 (2011) 013112.

\bibitem{ref:omelchenko13}
  O.E.~Omel'chenko,
  Coherence-incoherence in a ring of non-locally coupled phase oscillator,
  Nonlinearity 26 (2013) 2469.

\bibitem{ref:kawamura07}
  Y.~Kawamura, H.~Nakao, Y.~Kuramoto,
  Noise-induced turbulence in nonlocally coupled oscillators,
  Phys. Rev. E 75 (2007) 036209.
  [arXiv:nlin/0702042]

\bibitem{ref:kawamura08}
  Y.~Kawamura, H.~Nakao, K.~Arai, H.~Kori, Y.~Kuramoto,
  Collective phase sensitivity,
  Phys. Rev. Lett. 101 (2008) 024101.
  [arXiv:0807.1285]

\bibitem{ref:kawamura10a}
  Y.~Kawamura, H.~Nakao, K.~Arai, H.~Kori, Y.~Kuramoto,
  Phase synchronization between collective rhythms of globally coupled oscillator groups: Noisy identical case,
  Chaos 20 (2010) 043109.
  [arXiv:1007.4382]

\bibitem{ref:kawamura10b}
  Y.~Kawamura, H.~Nakao, K.~Arai, H.~Kori, Y.~Kuramoto,
  Phase synchronization between collective rhythms of globally coupled oscillator groups: Noiseless nonidentical case,
  Chaos 20 (2010) 043110.
  [arXiv:1007.5161]


\bibitem{ref:kawamura11}
  Y.~Kawamura, H.~Nakao, Y.~Kuramoto,
  Collective phase description of globally coupled excitable elements,
  Phys. Rev. E 84 (2011) 046211.
  [arXiv:1110.0914]




\bibitem{ref:nakao12}
  H.~Nakao, T.~Yanagita, Y.~Kawamura,
  Phase description of stable limit-cycle solutions in reaction-diffusion systems,
  Procedia IUTAM 5 (2012) 227-233.


\bibitem{ref:kori09}
  H.~Kori, Y.~Kawamura, H.~Nakao, K.~Arai, Y.~Kuramoto,
  Collective-phase description of coupled oscillators with general network structure,
  Phys. Rev. E 80 (2009) 036207.


\bibitem{ref:masuda09}
  N.~Masuda, Y.~Kawamura, H.~Kori,
  Analysis of relative influence of nodes in directed networks,
  Phys. Rev. E 80 (2009) 046114.

\bibitem{ref:masuda10}
  N.~Masuda, Y.~Kawamura, H.~Kori,
  Collective fluctuations in networks of noisy components,
  New J. Phys. 12 (2010) 093007.

\bibitem{ref:kori12}
  H.~Kori, Y.~Kawamura, N.~Masuda,
  Structure of cell networks critically determines oscillation regularity,
  J. Theor. Biol. 297 (2012) 61-72.


\bibitem{ref:jalics10}
  G.B.~Ermentrout, J.Z.~Jalics, J.E.~Rubin,
  Stimulus-driven traveling solutions in continuum neuronal models with a general smooth firing rate function,
  SIAM J. Appl. Math. 70 (2010) 3039-3064.

\bibitem{ref:kilpatrick12}
  Z.P.~Kilpatrick, G.B.~Ermentrout,
  Response of traveling waves to transient inputs in neural fields,
  Phys. Rev. E 85 (2012) 021910.

\bibitem{ref:lober12}
  J.~L\"ober, M.~B\"ar, H.~Engel,
  Front propagation in one-dimensional spatially periodic bistable media,
  Phys. Rev. E 86 (2012) 066210.

\bibitem{ref:ko09}
  T.-W.~Ko, G.B.~Ermentrout,
  Phase response curves of coupled oscillators,
  Phys. Rev. E 79 (2009) 016211.

\bibitem{ref:toenjes09}
  R.~T\"onjes, B.~Blasius,
  Perturbation analysis of complete synchronization in networks of phase oscillators,
  Phys. Rev. E 80 (2009) 026202.

\bibitem{ref:cross12}
  M.C.~Cross,
  Improving the frequency precision of oscillators by synchronization,
  Phys. Rev. E 85 (2012) 046214.

\bibitem{ref:cross13}
  J.-M.A.~Allen, M.C.~Cross,
  Frequency precision of two-dimensional lattices of coupled oscillators with spiral patterns,
  Phys. Rev. E 87 (2013) 052902.

\bibitem{ref:risken89}
  H.~Risken,
  The Fokker-Planck Equation: Methods of Solution and Applications,
  Springer, New York, 1989.

\bibitem{ref:gardiner97}
  C.W.~Gardiner,
  Handbook of Stochastic Methods: For Physics, Chemistry and the Natural Sciences,
  Springer, New York, 1997.

\bibitem{ref:yoshimura08}
  K.~Yoshimura, K.~Arai,
  Phase reduction of stochastic limit cycle oscillators,
  Phys. Rev. Lett. 101 (2008) 154101.

\bibitem{ref:teramae09}
  J.N.~Teramae, H.~Nakao, G.~B.~Ermentrout,
  Stochastic phase reduction for a general class of noisy limit cycle oscillators,
  Phys. Rev. Lett. 102 (2009) 194102.

\bibitem{ref:nakao10}
  H.~Nakao, J.N.~Teramae, D.S.~Goldobin, Y.~Kuramoto,
  Effective long-time phase dynamics of limit-cycle oscillators driven by weak colored noise,
  Chaos 20 (2010) 033126.

\bibitem{ref:goldobin10}
  D.S.~Goldobin, J.N.~Teramae, H.~Nakao, G.B.~Ermentrout,
  Dynamics of limit-cycle oscillators subject to general noise,
  Phys. Rev. Lett. 105 (2010) 154101.

\bibitem{ref:teramae04}
  J.N.~Teramae, D.~Tanaka,
  Robustness of the noise-induced phase synchronization in a general class of limit cycle oscillators,
  Phys. Rev. Lett. 93 (2004) 204103.

\bibitem{ref:goldobin05}
  D.S.~Goldobin, A.~Pikovsky,
  Synchronization of self-sustained oscillators by common white noise,
  Physica A 351 (2005) 126-132.

\bibitem{ref:nakao07}
  H.~Nakao, K.~Arai, Y.~Kawamura,
  Noise-induced synchronization and clustering in ensembles of uncoupled limit-cycle oscillators,
  Phys. Rev. Lett. 98 (2007) 184101.

\bibitem{ref:kurebayashi12}
  W.~Kurebayashi, K.~Fujiwara, T.~Ikeguchi,
  Colored noise induces synchronization of limit cycle oscillators,
  Europhys. Lett. 97 (2012) 50009.

\bibitem{ref:haim92}
  D.~Haim, O.~Lev, L.M.~Pismen, M.~Sheintuch,
  Modeling periodic and chaotic dynamics in anodic nickel dissolution,
  J. Phys. Chem. 96 (1992) 2676-2681.

\bibitem{ref:kiss08}
  I.~Z.~Kiss, Y.~Zhai, J.L.~Hudson,
  Resonance clustering in globally coupled electrochemical oscillators with external forcing,
  Phys. Rev. E 77 (2008) 046204.

\end{thebibliography}
\end{document}